\documentclass[12pt]{article}

\usepackage{epsfig}
\usepackage{graphicx}
\usepackage{amsmath}
\usepackage{amssymb}
\usepackage{latexsym}
\usepackage{color}
\usepackage{mathrsfs}
\usepackage{natbib}
\usepackage{hyperref}
\usepackage{appendix}
\usepackage[figuresright]{rotating}
\usepackage{dsfont}
\usepackage{longtable}
\graphicspath{{./figures/}}
\usepackage{amsfonts,epsfig,eurosym,color}
\usepackage[latin1]{inputenc}
\usepackage{amsmath,amssymb,amsfonts, amsxtra}
\usepackage{graphicx,psfrag}
\usepackage{hyperref}
\usepackage{enumerate}
\RequirePackage[caption=false]{subfig}
\usepackage{dsfont}
\usepackage{rotating}
\usepackage{url}
\usepackage{pdfsync}
\usepackage{colortbl}
\usepackage{natbib}
\usepackage[UKenglish]{babel}
\usepackage{natbib}
\usepackage{epstopdf}
\usepackage[normalem]{ulem}
\usepackage{multirow}
\usepackage{bm}

\newcommand{\e}{\mathbb{E}}
\newcommand{\var}{\mathbb{V}\text{ar}}

\DeclareMathOperator{\nach}{ne}
\DeclareMathOperator{\T}{^{\mathsf T}}

\makeatletter
\newcommand*\bdot{\mathpalette\bdot@{.7}}
\newcommand*\bdot@[2]{\mathbin{\vcenter{\hbox{\scalebox{#2}{$\m@th#1\bullet$}}}}}
\makeatother

\def\1{\mathds{1}}
\def\e{\mathds{E}}

\def\bSig\mathbf{\Sigma}

\usepackage[doublespacing]{setspace}

\begin{document}

\def\spacingset#1{\renewcommand{\baselinestretch}%
{#1}\small\normalsize} \spacingset{1}


  \title{\bf  Partial and semi-partial measures of spatial associations for multivariate lattice data}
  \author{Matthias Eckardt\thanks{
    The authors gratefully acknowledge \textit{please remember to list all relevant funding sources in the unblinded version}}\hspace{.2cm}\\
    Department of Mathematics, University Jaume I, Castell\'{o}n, Spain\\
    and \\
   Jorge Mateu\\
    Department of Mathematics, University Jaume I, Castell\'{o}n, Spain}
  \maketitle

\bigskip
\begin{abstract}
This paper is concerned with the development of partial and semi-partial measures of spatial associations in the context of multivariate spatial lattice data which describe the global or local associations among spatially aggregated measurements for pairs of different components conditional on all remaining components. The new measures are illustrated using aggregated data on crime counts at ward level.
\end{abstract}

\noindent%

\spacingset{1.45} 

\begin{abstract}
This paper concerns the development of partial and semi-partial measures of spatial associations in the context of multivariate spatial lattice data which describe global or local associations among spatially aggregated measurements for pairs of different components conditional on all remaining components. The new measures are illustrated using aggregated data on crime counts at ward level.
\textit{Geary's C,  LISA,  Moran's I, multivariate areal data, partial hotspot detection}
\end{abstract}

\section{Introduction}

Spatial areal data is a particular important field of research which is, in the most simplest form, characterised by a set of spatially aggregated measurements for one outcome of interest recorded over a countable collection of disjoint spatial units - commonly denoted as sites. Examples include aggregated data on  incidence, prevalence and mortality rates or crime counts per areal unit.  Usually, the sites are artificial by nature due to administrative reasons and obey an underlying planar interconnection structure which could be used to calculate spatially lagged values, e.g. the average outcome computed over all spatial neighbours of a particular site, which in turn allow to extend the well-established toolbox of time series models and specifications to the spatial domain. 

For the univariate case, a wide range of different spatial models and specifications can be found in the literature. Apart from spatial autocorrelation specifications such as the conditional autoregressive \citep{Besag1972,Besag1974a} or the  simultaneous autoregressive \citep{Whittle1954} model, different global and local measures of spatial autocorrelation have been proposed aiming to detect and measure spatial clustering. While the global measures such as Moran's $I$ \citep{moran}, Geary's contiguity ratio $C$ \citep{geary} or Getis' and Ord's $G$ and $G^\star$ \citep{geatisord} yield a quantification of the overall spatial clustering of the outcome, local measures including the work of \cite{Anselin1995}, \cite{getisfranklin87}, \cite{getisfranklin10}, \cite{geatisord}, and \cite{ordgetis12} focus on the local variation in autocorrelation by taking each sites' relative contribution on the overall autocorrelation into account (see \cite{Banerjee2004}, \cite{Cressie1993},  \cite{Forth97},   \cite{griffith1987}, \cite{llyod2011}, \cite{odland1988spatial},  \cite{OrdGetis1995} and \cite{Ripley1981} for a general overview).

In view of the rapid development of geoinformation systems and storage capacities, and open-source data supply, the availability and accessibility of immense areal data strongly increases, and feasible data analysis tools for pattern recognition and feature detection  have gained considerable importance in a myriad of different disciplines. Unlike the classical univariate case, modern (open-source) databases on areal data provide information on various different outcomes in space, each of which is recorded over a congruent set of $n$ interconnected spatial units, which might be affected by potential (lagged) associations within and between different  outcomes.  One challenge for spatial methods development is therefore not only to extract relevant information from possibly high-dimensional data on multivariate associations but also to control for potential effects of alternative components on the spatial association among pairs of spatial patterns.

However, although various well-established methods exist for the investigation and quantification of spatial autocorrelations, only a limited number of papers have considered multivariate spatial autocorrelation statistics. Focussing on the global Moran's $I$ statistic, a first treatment of multivariate autocorrelations 
was presented by \cite{Wartenberg1985} who extended the principal component framework to the spatial case, leading to  the multivariate spatial analysis based on Moran's I  
  (MULTISPATI) framework of \cite{dray2008}  which  introduces a row-sum standardised spatial weight matrix in the statistical triplet notation. A bivariate extension of the Moran scatterplot was discussed by \cite{Anselin02visualizingmultivariate} which depicts the lagged value of the first variable on the vertical and the original value of the second variable on the horizontal axis. 
A different measure for bivariate spatial associations which is similar in spirit but different in detail is the so-called $L$ statistic, which integrates Pearson's $r$ statistic and Moran's $I$ as proposed by \cite{lee2001}. A discussion of different  analysis techniques for multivariate areal data was presented by  \cite{friendly2007} and \cite{dray2011} and applied to  Guerry's data \citep{guerry1833} on the moral statistics in France (see  \cite{guerrydata}). Finally, apart from the above global measure, \cite{Anselin2019+} presented a multivariate extension of the local Geary's contiguity ratio $C$ defined as a weighted mean of the squared distances in the multivariate attribute space between the value observed at a given site and those at its spatial neighbours. While these global and local approaches allow to investigate clustering and regional variation of autocorrelation in the presence of multiple outcomes, they do not provide information on cross-associations between pairs of different outcomes under control for any alternative outcome. To this purpose, we concern here with the development of partial and semi-partial measures of spatial association for multivariate spatial areal data which describe the spatial association between pairs of different outcomes conditional on all alternative outcomes over a congruent spatial area. 
 
The remainder of this paper is structured as follows. Section 2 briefly recapitulates the concept of spatial proximity, reviews some univariate global and local measures of spatial association and discusses extensions to the bivariate case. These results will then be used to derive the partial and semi-partial measures of spatial association. An application to multivariate data on  aggregated crime counts for three different types of offences is covered in Section 3. The paper ends with some conclusions in Section 4.

\section{Global and local measures of spatial association}

This section develops the idea of partial and semi-partial measures for multivariate spatial areal data of the form $\lbrace \mathbf{x}(\mathbf{s}_n)\rbrace =\lbrace (x_1(\mathbf{s}_n), \ldots  x_d(\mathbf{s}_n))\rbrace\T$ which is understood as a realisation of a $d$-variate spatial areal process $\lbrace X(\mathbf{s}_n)\rbrace$ on $\mathbf{S}_L\subseteq \mathds{Z}^2$ with $(\mathbf{s}_1,\ldots,\mathbf{s}_n)$ denoting the locations in the lattice $\mathbf{S}_L$. The stochastic process $\lbrace \mathbf{X}(\mathbf{s}_n)\rbrace$ is formed by $d$ components denoted by $X_i(\mathbf{s}_n)$.

\subsection{Spatial proximity of lattice entities}\label{sec:proximity}

To discuss the framework of spatial autocorrelation measures, the spatial proximity matrix $\mathbf{W}$, which plays a fundamental role in the classical analysis of spatial areal processes, has to be presented first. Several authors have contributed to this field and a large body of research  exists on the specification of the spatial proximity matrix. For a detailed treatment of different conceptual definitions of $\mathbf{W}$  we refer the interested reader to  \cite{Banerjee2004} and \cite{Cressie1993} and the references therein.    

In general, the proximity matrix can be specified directly from the spatial lattice and is primarily used to define a planar neighbourhood structure over $\mathbf{S}_L$ connecting spatially close sites. For each site $\mathbf{s}_i$ contained in $\mathbf{S}_L$, the spatial neighbours  $\nach_s(\mathbf{s}_i)$ of $\mathbf{s}_i$ are defined as all alternative sites $\mathbf{s}_j$ which are said to be spatially close to $\mathbf{s}_i$ with respect to a given criterion. The information on neighbouring sites is then captured in the proximity matrix such that, 
in the most general case, the $ij$-th element $w_{ij}$ of $\mathbf{W}$ is defined by
\[
w_{ij}=
\begin{cases}
1\text{~if~} \mathbf{s}_j\in\nach_s(\mathbf{s}_i)\\
0\text{~otherwise}
\end{cases}
\]
where $w_{ij}=w_{ji}$ due to the symmetry of $\mathbf{W}$. Notice that, by convention, $w_{ii}=0$ as $\mathbf{s}_i\notin\nach_s(\mathbf{s}_i)$.
Hence, in mathematical parlance, the spatial neighbourhood structure is a undirected graph representing the spatial lag structure among the sites of $\mathbf{S}_L$.  Apart from this binary coding, the proximity matrix could also display numerical information such as the pairwise reciprocal intercentroidal distance. However, for simplicity, these alternative specifications will not be covered here. 

We now briefly review the specification of the proximity matrix $\mathbf{W}$ through commonly shared borders and intercentroidal distances between different lattice entities. Under the commonly shared borders specification, two sites are said to be neighbours if both (lattice) entities share a common border. Apart from this (first-order) neighbourhood specification, and extensions to higher-order neighbourhoods could easily be implemented. For example, the set of second-order neighbours of site $\mathbf{s}_i$ could be defined as all  sites which share a common border with the first-order neighbours except $\mathbf{s}_i$. 

Another criterion for the specification of the neighbourhood structure relies on the thresholding of intercentroidal distances. Under this approach, the neighbouring sites $\mathbf{s}_j$ of $\mathbf{s}_i$ are defined as the set of all those sites whose intercentroidal distances $d_L(\mathbf{s}_i,\mathbf{s}_j)$ are below or equal to a prespecified threshold distance $\varsigma
$. Another approach would be to consider only the $k$-nearest neighbouring sites $\mathbf{s}_j$ of $\mathbf{s}_i$ which satisfy $d_L(\mathbf{s}_i,\mathbf{s}_j)\leq\varsigma
$. Alternatively, one could also consider distance bins, say $(d_{L}(\cdot)\geq \varsigma
_1, d_{L}(\cdot)<\varsigma
_2)$, such that the neighbouring sites $\mathbf{s}_j$ of $\mathbf{s}_i$ are defined as those spatial units whose intercentroidal distances are greater than or equal to the first threshold $\varsigma
_1$ and below the value of the second threshold $\varsigma_2$.   
 
\subsection{The univariate case}

This section describes two popular global measures commonly used to investigate global associations among the measurements over spatially lagged sites, namely Geary's contiguity ratio $C$ \citep{geary} and Moran's $I$ \citep{moran}. Both statistics can be understood as spatial analogues of classical time series statistics and reflect the strength of spatial autocorrelations of the variables among the spatial entities. Besides these global measures, a localised version of Moran's $I$ statistic (the local Moran's $I_{i}$) is also covered. Different from global measures, local measures of spatial association, which have been coined local indicators of spatial association (LISA) by \citet{Anselin1995}, reveal information on the influence of each single site on the global measure and also indicate substantial local deviations from Gaussian white noise. In words, LISA functions decompose the information provided by global measures into local information on the contribution of each site on the spatial autocorrelation.   

First, the global statistics are presented. Geary's $C$ is defined by  
\begin{equation}\label{eq:geary}
C = \frac{n-1}{2\sum_i\sum_jw_{ij}}\frac{\sum_i\sum_jw_{ij}(X(\mathbf{s}_i)-X(\mathbf{s}_j))^2}{\sum_i(X(\mathbf{s}_i)-\mu_X)^2}
\end{equation}
where $n$ is the number of sites under study, $\mu_X=\e\left[\lbrace X(\mathbf{s}_n)\rbrace\right]$ and $w_{ij}$ is the $ij$-th element of $\mathbf{W}$. We note that $C$ is non-negative and ranges between $0$ and $2$. Under the null hypothesis of no spatial autocorrelation among neighbouring sites, this measure has the expected value of one and could be interpreted as follows (see \cite{geary} and \cite{griffith1987} for proof). An estimate $\widehat{C}$ of $C$ can be calculated from empirical data using the formula
\begin{equation*}
\widehat{C}=\frac{\sum_i\sum_jw_{ij}(x(\mathbf{s}_i)-x(\mathbf{s}_j)^2}{2\sum_i\sum_jw_{ij}\widehat{\sigma^2_C}}
\end{equation*} 
where $\widehat{\sigma^2_C}=\sum_i\sum_j(x(\mathbf{s}_i)-\widehat{\mu_X})^2/(n-1)$ is the sample variance with $\widehat{\mu_X}=1/n\sum_i(x(\mathbf{s}_i)$. If $\e\left[C\right]> \widehat{C}$, similar values have been recorded over  neighbouring sites, while the presence of dissimilar values among  neighbouring sites is indicated by $\e\left[C\right]< \widehat{C}$. Hence, values of the empirical ratio $\widehat{C}$ between $0$ and $1$ reflect positive autocorrelation. 

The second global measure, Moran's $I$, is defined by 
\begin{equation*}
I= \frac{n}{\sum_i\sum_jw_{ij}}\frac{\sum_i\sum_jw_{ij}(X(\mathbf{s}_i)-\mu_X)(X(\mathbf{s}_j)-\mu_X)}{\sum_i(X(\mathbf{s}_i)-\mu_X)^2}. 
\end{equation*}
Different from Geary's $C$, Moran's $I$ takes values between $-1$ and $1$.
Under the null hypothesis of no spatial autocorrelation, this measure is approximately Gaussian distributed with mean $\e\left[I \right]=-1/(n-1)$ and variance 
\begin{equation*}
\var\left[I\right]=\frac{n^2(n-1)\frac{1}{2}\sum_{i\neq j}(w_{ij}+w_{ji})^2-n(n-1)\sum_k(\sum_jw_{kj}+\sum_iw_{ik})^2-2(\sum_{i\neq j}w_{ij})^2}{(n+1)(n-1)^2(\sum_{i\neq j}w_{ij})^2}
\end{equation*}
(cf. \cite{Banerjee2004}). Obviously,  $\e\left[I\right]$ is essentially zero for large $n$. Similar to Geary $C$ statistic, an estimate $\widehat{I}$ of $I$ can be calculated from empirical data using the formula
\begin{equation*}
\widehat{I}=\frac{\sum_i\sum_jw_{ij}(x(\mathbf{s}_i)-\widehat{\mu_X})(x(\mathbf{s}_j)-\widehat{\mu_X})}{\widehat{\sigma^2_I}\sum_i\sum_jw_{ij}}
\end{equation*}
where $\widehat{\sigma^2_I}=\sum_i\sum_jx(\mathbf{s}_i)-\widehat{\mu_X})^2/n$ is the sample variance (see \cite{HKUHUB_10722_192432}). The interpretation of Moran's $I$ is opposed to Geary's $C$ such that positive autocorrelation among neighbouring sites exists if $\e\left[I\right]< \widehat{I}$.

Lastly, the local Moran's $I_{i}$ statistic is presented. For the $i$-th site of $\mathbf{S}_L$, this local measure of spatial association is defined by
\begin{equation*}
I_{i}=\frac{(X(\mathbf{s}_i)-\mu_X)}{\sum_{k=1}^{n}(X(\mathbf{s}_k)-\mu_X)^2/(n-1)}\sum_{j=1}^{n}w_{ij}(X(\mathbf{s}_j)-\mu_X)
\end{equation*}
and can be estimated by 
\begin{equation*}
\widehat{I_{i}}=\frac{x(\mathbf{s}_i)-\widehat{\mu_X}}{\widehat{\sigma^2_X}}\sum_{j=1}^{n}w_{ij}x(\mathbf{s}_j)
\end{equation*}
where $\widehat{\sigma^2_X}=\sum_{i=1}^n(x(\mathbf{s}_i)-\widehat{\mu_X})^2/n$ is the sample variance (see \cite{Anselin1995}).
The interpretation of the local Moran's $I_{i}$ statistic derives directly from the interpretation of Moran's $I$. For example, if $\e\left[I_{i}\right]< \widehat{I}_{i}$ with $\e\left[I_{i}\right]=-\sum_iw_{ij}/(n-1)$ it follows that the neighbouring sites of $\mathbf{s}_i$ are similar in value to the measurement made at $\mathbf{s}_i$.  

\subsection{Multivariate measures of spatial association}

While the previous section has recapitulated the univariate case where only one outcome is observed over a set of interconnected areal units, we now focus on extensions of the above measures to $d$-variate areal processes where several different outcomes are observed over a congruent spatial lattice $\mathbf{S}_L$. For simplicity, the bivariate case is considered first.

For two different components $X_i(\mathbf{s}_n)$ and $X_j(\mathbf{s}_n)$ of $\mathbf{X}(\mathbf{s}_n)$, a  generalisation of Geary's global contiguity ration $C$  yields

\begin{equation*}
C_{ij} = \frac{n-1}{2\sum_i\sum_jw_{ij}}\frac{\sum_i\sum_jw_{ij}(X_i(\mathbf{s}_i)-X_j(\mathbf{s}_j))^2}{\sum_i( X_i(\mathbf{s}_i)-\mu_{X_i}) \times \sum_j (X_j(\mathbf{s}_j)-\mu_{X_j})}
\end{equation*}

where $\mu_{X_i}$ and $\mu_{X_j}$ are the first-order moments of $X_i(\mathbf{s}_n)$ and $X_j(\mathbf{s}_n)$, respectively. Different from  \eqref{eq:geary}, this expression provides information on the cross-correlation between two distinct outcomes, eg. how two components are clustered in space. However, we note that, unlike the univariate contiguity ration $C$, the bivariate version $C_{ij}$ is not symmetric any more such that $C_{ij}\neq C_{ji}$ in general. 

Likewise, for the global Moran's $I$, we define a bivariate cross-correlation index by 
\begin{equation*}
I_{ij} = \frac{n}{\sum_i\sum_jw_{ij}}\frac{\sum_i\sum_jw_{ij}(X_i(\mathbf{s}_i)-\mu_{X_i})(X_j(\mathbf{s}_j)-\mu_{X_j})}{\sum_i (X_i(\mathbf{s}_i)-\mu_{X_i}) \times \sum_j (X_j(\mathbf{s}_j)-\mu_{X_j})}. 
\end{equation*}

As for the bivariate $C_{ij}$, this expression is again not symmetric such that $I_{ij}\neq I_{ji}$. 

Aiming to the local variation between two distinct components, we define a bivariate local version of Moran's $I$ as
\begin{equation*}
I_{i,j}=\frac{(X_i(\mathbf{s}_i)-\mu_{X_i})}{\sum_{k=1}^{n}(X_i(\mathbf{s}_k)-\mu_{X_i})^2/(n-1)}\sum_{j=1}^{n}w_{ij}(X_j(\mathbf{s}_j)-\mu_{X_j}).
\end{equation*}

While the above equations provide information on the spatial cross-correlation between two different outcomes and allow to quantify the spatial variation across the area under study, they do not help to distinguish between direct and induced  cross-correlation potentially caused by alternative outcomes and might display spurious interrelations. For this reason, we now extend the above equations to the multivariate case and discuss partial and semi-partial versions of Geary's $C$ and Moran's $I$. To this end, let $X_{i|\mathbf{c}}(\mathbf{s}_n)$ and $X_{j|\mathbf{c}}(\mathbf{s}_n)$ denote the outcome of $X_i(\mathbf{s}_n)$ and $X_j(\mathbf{s}_n)$ conditional on the  component $X_{\mathbf{c}}(\mathbf{s}_n)$, respectively, where 
\begin{equation*}
X_{\mathbf{c}}(\mathbf{s}_n)=\lbrace (X_1(\mathbf{s}_n)\ldots, X_{i-1}(\mathbf{s}_n), X_{i+1}(\mathbf{s}_n),\ldots,X_{j-1}(\mathbf{s}_n),X_{j+1}(\mathbf{s}_n),\ldots, X_d(\mathbf{s}_n))\rbrace\T
\end{equation*}

are all components of $\mathbf{X}(\mathbf{s}_n)$ except $X_i(\mathbf{s}_n)$ and $X_j(\mathbf{s}_n)$. Likewise, let $\mu_{X_i|\mathbf{c}}$  and $\mu_{X_j|\mathbf{c}}$ denote the first-order moments of  $X_{i|\mathbf{c}}(\mathbf{s}_n)$ and $X_{j|\mathbf{c}}(\mathbf{s}_n)$.      

Then, we can define a partial version of Geary's contiguity ratio as  
  
\begin{equation}
\label{eq:partialC}
C_{ij|\mathbf{c}} = \frac{n-1}{2\sum_i\sum_jw_{ij}}\frac{\sum_i\sum_jw_{ij}(X_{i|\mathbf{c}}(\mathbf{s}_i)-X_{j|\mathbf{c}}(\mathbf{s}_j))^2}{\sum_i (X_{i|\mathbf{c}}(\mathbf{s}_i)-\mu_{X_i|\mathbf{c}}) \times \sum_j (X_{j|\mathbf{c}}(\mathbf{s}_j)-\mu_{X_j|\mathbf{c}})}
\end{equation}
which provides information on how the components $X_i(\mathbf{s}_n)$ and $X_j(\mathbf{s}_n)$ are clustered in space conditional on $X_{\mathbf{k}}(\mathbf{s}_n)$. Notice that, as for the bivariate version of Geary's  $C$, $C_{ij|\mathbf{c}}$ is again not symmetric such that $C_{ij|\mathbf{c}}\neq C_{ji|\mathbf{c}}$.

Similarly, a partial version of Moran's $I$ can defined as 
\begin{equation}
\label{eq:partialI}
I_{ij|\mathbf{c}} = \frac{n}{\sum_i\sum_jw_{ij}}\frac{\sum_i\sum_jw_{ij}(X_{i|\mathbf{c}}(\mathbf{s}_i)-\mu_{X_i|\mathbf{c}})(X_{j|\mathbf{c}}(\mathbf{s}_j)-\mu_{X_j})}{\sum_i X_{i|\mathbf{c}}(\mathbf{s}_i)-\mu_{X_i|\mathbf{c}} \times \sum_j X_{j|\mathbf{c}}(\mathbf{s}_j)-\mu_{X_j|\mathbf{c}}}.
\end{equation}

Apart from this global partial measure, a local partial Moran's $I$ can be constructed similarly to the bivariate case yielding 
\begin{equation*}
I_{i,j|\mathbf{c}}=\frac{(X_i|\mathbf{c}(\mathbf{s}_i)-\mu_{X_i|\mathbf{c}})}{\sum_{k=1}^{n}(X_{i|\mathbf{c}}(\mathbf{s}_k)-\mu_{X_{i|\mathbf{c}}})^2/(n-1)}\sum_{j=1}^{n}w_{ij}(X_{j|\mathbf{c}}(\mathbf{s}_j)-\mu_{X_{j|\mathbf{c}}}).
\end{equation*}

Besides, for the three components $X_i(\mathbf{s}_n), X_j(\mathbf{s}_n)$ and $X_k(\mathbf{s}_n)$ of $\mathbf{X}(\mathbf{s}_n)$ and using some well know results, \eqref{eq:partialC} and \eqref{eq:partialI} simplify to 
\begin{equation*}
C_{ij|k}=\frac{C_{ij} -C_{ik}C_{jk}}{\sqrt{1-C_{ik}^2}\sqrt{1-C_{jk}^2}} 
\end{equation*}
and
\begin{equation*}
I_{ij|k}=\frac{I_{ij} -I_{ik}I_{jk}}{\sqrt{1-I_{ik}^2}\sqrt{1-I_{jk}^2}}
\end{equation*}
such that the partial measures can directly be obtained from the bivariate expressions.

While \eqref{eq:partialC} and \eqref{eq:partialI} describe the cross-correlation of  $X_i(\mathbf{s}_n)$ and  $X_j(\mathbf{s}_n)$ conditional on all remaining components of $\mathbf{X}(\mathbf{s}_n)$, it is sometimes of interest to hold $X_k(\mathbf{s}_n)$ constant for just one component, say $X_i(\mathbf{s}_n)$. In that case,  one might consider the semi-partial versions of Geary 's $C$ and Moran's $I$ defined by
\begin{equation*}
C_{i.j|k}=\frac{C_{ij} -C_{ik}C_{kj}}{\sqrt{1-C_{jk}^2}} 
\end{equation*} 
and
\begin{equation*}
I_{i.j|k}=\frac{I_{ij} -I_{ik}I_{kj}}{\sqrt{1-I_{jk}^2}} ,
\end{equation*}
respectively.

\section{Application to aggregated crime data}

For illustration of the proposed partial statistics, we consider a subset of three different types of offences recorded over London within a one month period in December 2015. For simplicity, we only discuss the results obtained for the bivariate and partial local Moran's $I$ in detail. The sample data under study provides information on aggregated crime counts per ward for anti-social behaviour, criminal damage and arson, and violence and sexual behaviour and was constructed from an open-source multivariate point pattern on crimes provided under the Open Government Licence by the British Home Office for London. The original source has been downloaded from {\url{http://data.police.uk/data/}}  and  contains the Longitude and Latitude for $14$ pre-classified crime categories at street-level, either within a one mile radius of a single point or within a custom area of a street recorded by the Metropolitan Police.

To provide a first impression of the sampled aggregated data, different univariate summary and autocorrelation characteristics calculated from all three types of crimes are reported in Table . Inspecting both spatial autocorrelation statistics, all three types of crime included showed a tendency towards positive autocorrelation.  

%

This impression is also supported considering global bivariate and partial $I$ statistics. 

\begin{figure}[bt]
\centering
\includegraphics[width=13.5cm]{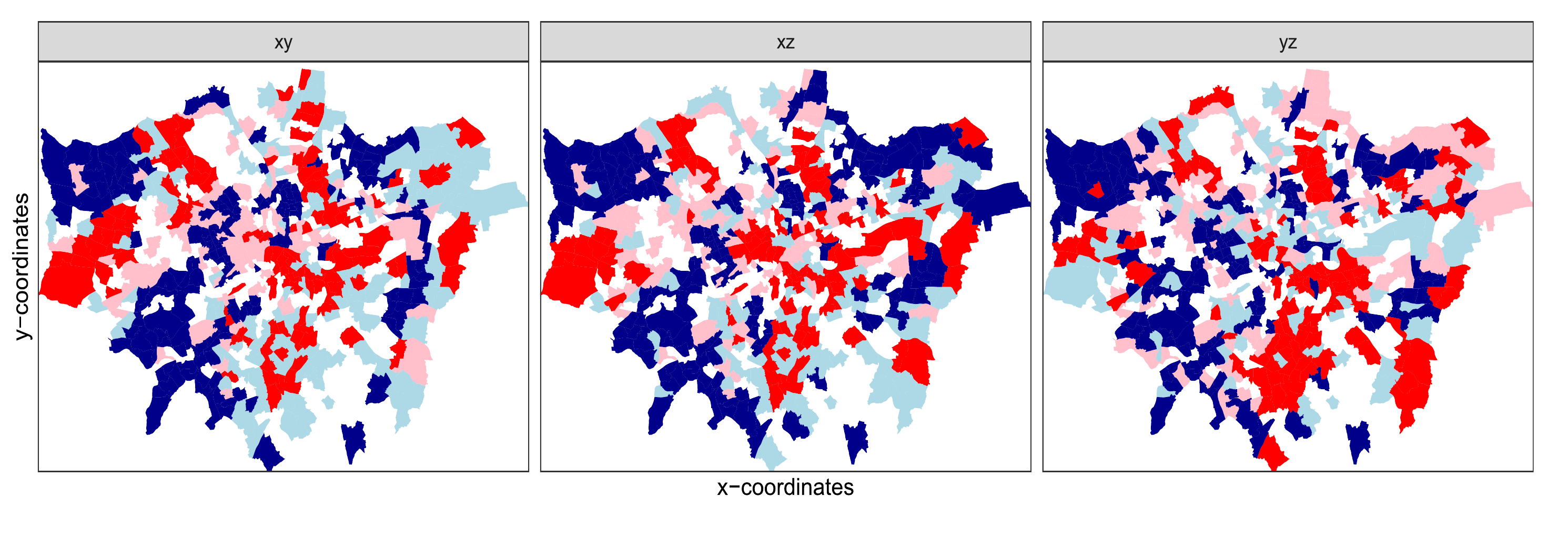}
\caption{Bivariate Moran's $I$ significance map at ward level computed from the aggregated crime counts. $x$ = Anti-social behaviour, $y$ = violence and sexual offence, and $z$ = criminal damage and arson. Colours are coded as follows: red = high-high, pink = high-low, light-blue = low-high and dark-blue = low-low. \label{fig:bivar}}
\end{figure}

Next, to investigate the spatial variations among the three types of crime included from a local perspective, bivariate and partial Moran's I significance maps were calculated.  Inference was carried out using a permutation approach to detect significant local variation at a $5$ percent level. The results of the bivariate Moran's $I$ statistic are depicted in  Figure \ref{fig:bivar}. Inspecting this plot, a clear concentration of local coldspot on the left sides of each subplot can be identified for all three pairings of crimes indicating that low crime counts of the first crime at theses areas are surrounded by low crime counts of the second crime. Besides, large hotspot areas can be detected. However, while we found some similarities between the spatial locations of coldspots among all three subplots, a clear distinction can be made between the locations of bivariate hotspots of anti-social behaviour and violence and sexual offences (left panel), anti-social behaviour and criminal damage and arson (middle panel) on the one hand and violence and sexual offence and criminal damage and arson (right panel) on the other hand.

\begin{figure}[bt]
\centering
\includegraphics[width=13.5cm]{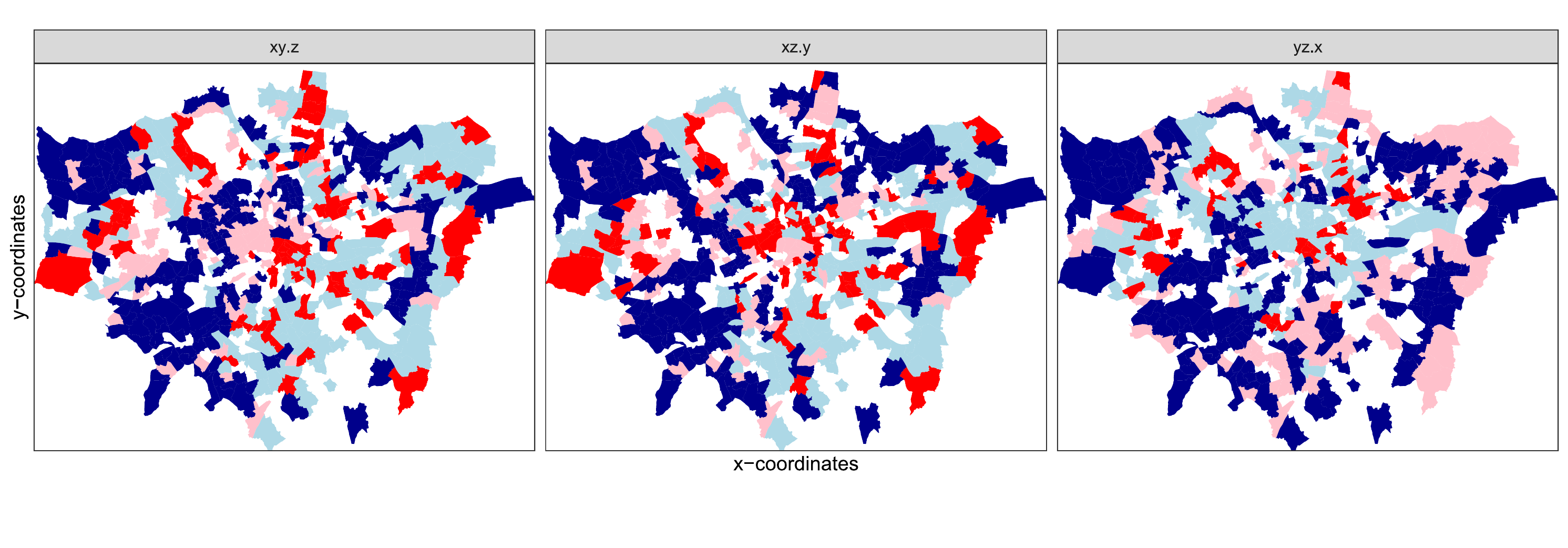}
\caption{Partial Moran's $I$ significance map at ward level computed from the aggregated crime counts. $x$ = Anti-social behaviour, $y$ = violence and sexual offence, and $z$ = criminal damage and arson. Colours are coded as follows: red = high-high, pink = high-low, light-blue = low-high and dark-blue = low-low. \label{fig:partail}}
\end{figure}

Turning to the partial Moran's $I$ significance maps shown in  
 Figure \ref{fig:partail}, a clear variation of hot- and coldspot patterns can be detected between the bivariate and partial significance maps. For the partial hotspot patterns, further coldspot regions are shown and appear in addition to those areas which were already classified as coldspot by the bivariate Moran's $I$ maps. In particular, a strong increase of coldspot regions on the right sides of all three panels is depicted. Further, while some clear bivariate hotspots are displayed in Figure \ref{fig:bivar}, only a few areas appeared as partial hotspots for all three pairings under study. In particular, most bivariate hotspot locations changed to areas of dissimilar pairwise local associations conditional on the alternative components. These areas indicated by pink and light blue suggest that high (resp. low) crime counts are surrounded by low (resp. high) neighbouring crime counts which might reveal potential spurious hotspot associations from a purely bivariate perspective.      

\section{Conclusions}

This paper has introduced a new class of partial and semi-partial measures of spatial
autocorrelation for the analysis of multivariate spatial areal data. These new introduced measures can describe both 
global or local associations among spatially aggregated measurements for pairs of different components conditional on all remaining ones. Different from global measures, these partial measures are able to reveal potential spurious associations from a purely bivariate or multivariate perspective. Indeed, partial global and local approaches allow to investigate clustering and regional variation of
autocorrelation in the presence of multiple outcomes, providing information on
cross-associations between pairs of different outcomes under control for any alternative outcome.

We have not considered the case where we have a multivariate case based on mixture of types of data, for example when some data comes on a lattice support, and other components are point patterns on the network. This is coming a common case, and would be natural and timely to provide the right framework to deal with new formats of spatial data.

\bibliographystyle{ecta}
\bibliography{partialMoran}

\end{document}